\documentclass[aps,pre,twocolumn]{revtex4}
\usepackage{graphics,amsmath}
\begin{document}
\title{Influence of external information in the minority game}
\author{M. A. R. de Cara}
\thanks{Current address: Adaptive Dynamics Network.
International Institute for Applied Systems Analysis.
Schlossplatz 1. A-2361 Laxenburg. Austria.}
\author{F. Guinea} 
\affiliation{
Instituto de Ciencia de Materiales de Madrid\\
Consejo Superior de Investigaciones Cient{\'\i}ficas\\
Cantoblanco, E-28049 Madrid, Spain}

\begin{abstract}
The influence of a fixed number of agents with the same fixed behavior
on the dynamics of the minority game is studied. Alternatively,
the system studied can be considered the minority game with a change
in the comfort threshold away from half filling. Agents in the frustrated, non
ergodic phase tend to overreact to the information provided by the 
fixed agents, leading not only to large fluctuations, but to deviations
of the average occupancies from their optimal values. Agents which
discount their impact on the market, or which use individual strategies
reach equilibrium states, which, unlike in the absence
of the external information provided by
the fixed agents, do not give the highest payoff
to the collective.
\end{abstract}
\pacs{89.65.-s , 89.65.Gh , 89.20.-a}
\date{\today}
\maketitle

\section{Introduction\label{s:intro}}
The minority game has become an extensively used model of some aspects
of financial markets~\cite{CZ97}.  It shows that complex behavior
can arise from relatively simple mathematical rules, used to define
a system of interacting agents. In addition, it is amenable to
analytical treatment~\cite{CMZ00}, and shows the usefulness of
the methods of statistical physics for the study of problems
of interest in economics, sociology, or biology~\cite{MG}.
The model has been extensively analyzed, and it shows
a phase transition between
an ergodic phase, where the agents reach a well defined stationary
state, and a non ergodic phase, where the evolution is strongly
dependent on the initial conditions~\cite{SMR99}. 
The ergodic phase can be well characterized by means of
the replica formalism, well known in studies of 
systems with quenched disorder. The disorder in the minority game
arises from random differences between the agents, associated
to the strategies at their disposal (see below). 
There is no similar degree
of understanding of the behavior of the agents in the non 
ergodic phase, where frustration and herding effects
play a major role in determining the long time
evolution. Relatively simple modifications of the rules of
the game change significantly the results, for the parameter
range where the ergodic phase occurs. These changes can modify,
or even suppress,
herding behavior. We can mention, among other variations,
evolution based 
schemes, which allow for the use of the opposite outcome predicted
by the ``best'' strategy~\cite{Jetal99}, agents which discount
the effect of their own choices on the market~\cite{MC01}, or agents
which use individual, instead of global, information~\cite{CPG00}. 
The case where agents discount their impact on the market can
be studied analytically, and it can be shown that the dynamics lead to
a stationary state with small volatitlity, and which
optimizes the benefits to the collective~\cite{MC01,CMZ00b,MM01}. 
It is known that, for the non ergodic phase 
in the standard version of the minority game, the available
information is arbitraged away, making the difference between 
the actual histories and random data irrelevant~\cite{C99}. 
On the other hand, the outcome of the game shows a significant
dependence on the initial bias in the scores of the strategies, when
this bias is allowed to have a finite value\cite{HC01}.

In the present work, we will analyze further how the external information
is processed in the non ergodic phase. For that purpose, we
will assume that a given number of agents make always the same choice,
inducing a bias in the outcome. The predictable behavior
of these fixed agents can be considered as an external information
source which can be processed by the remaining active agents.
If the active agents were playing
at random, the minority group would tend to be the one not preferred by
the fixed agents.
This situation corresponds to having a given number
of correlated producers, in the generalization of the minority game
described in~\cite{CMZ00b,MM01}.  Alternatively, we can consider that
the ``comfort threshold'' for the active agents has been
shifted away from half filling by the presence of the agents with
fixed choices. This situation was already considered in the
initial version of the minority game~\cite{A94}, and it
has been further studied in~\cite{Jetal99b}. An extension of the
analytical results for the standard minority game to a situation
where the ``comfort threshold'' has been shifted
can be found in~\cite{CMO03}. A situation where all strategies
used by the agents are biased towards a given outcome
is discussed in~\cite{Yetal03}. A related situation is that in which
some agents prefer to be in the majority, considered in~\cite{MGM03}.
The existence of these ``trend followers'', however, is
not a source of information for the other agents.

The models studied will be more precisely defined in the following section.
We present the main results in section III. 
Finally, section IV summarizes the conclusions,
and compares the results with related work.
\section{The models\label{s:models}}
We study the minority game defined in the usual way. There are $N$ agents
which use $s$ strategies each,
assigned initially at random. These strategies
associate a given binary outcome to a series of $m$ binary numbers, which
represent the history of the game in the previous $m$ time steps.
The number of possible strategies is $2^{2^m}$. The goal of the agents
is to choose the minority group, that is, the one chosen by less
than half of the agents, $N/2$.  There is a given number of agents
$N_f$ which always make the same choice, 1. 
Hence, the number of ``active'' agents is $N - N_f$.
The maximum number of active agents
which can win at a given time is, obviously, $N/2$, when $N/2 - N_f$
active agents make the choice 1, and the remaining $N/2$ choose 0. 
The game becomes trivial if $N_f \ge N/2$, as all the active agents
will profit from making the choice 0. Note that in the standard
version of the minority game there will be,
on the average, a fraction $2^{-s}$ of agents unable to make this choice,
as the strategies available to them  lead only to choice 1. The results to be 
discussed are averages over the possible distributions of strategies
among agents.

We study three versions of the model, 
which differ in the way the score of the
strategies available to the agents are updated, or in the information
processed by the agents:

i) The standard minority game, as defined in~\cite{CZ97}. Each agent updates
the score of the strategies available to it according to whether the
predicted outcome was successful (one point is added to the score) 
or unsuccessful (zero points are added). 

ii) The individual minority game,
as defined in~\cite{CPG00}. The input used by each agent in order
to decide the outcome predicted by a given strategy is the 
succession of events that it has experienced. A given (individual) history thus
corresponds to the series of choices made by the agent. 

iii) The 
minority game where agents discount the impact of the strategy which
they have used on the global result~~\cite{CMZ00b,MC01,CMZ00}.
In order to take the impact into account, the score of the strategies
is updated considering what would have happened if the agent
had taken the opposite decision, and rewards
the strategy used. For that purpose, we follow
the linear payoff introduced in ref.~~\cite{CMZ00},
which considers an increase in the score of each strategy $s$
of agent $i$ in time $t$ of
$\Delta= - a_{s,i}(t) A(t)/P + \eta\delta_{s,s_i(t)}/P$;
$a_{s,i}$ is the prediction of strategy $s$ (in terms of $\{-1,1\}$),
$A(t) = N_1(t) - N_0(t)$, and $\eta$ is the reward to the strategy
played. In the following we use  $\eta=0.5$.

In all three cases, the total number of agents is $N$, 
which is taken to be an odd number. Then,
the number of winners cannot exceed $( N - 1 ) / 2$. 
The fixed agents can also be on the winning
side, and the same bound also applies to the winners among
the active agents. In all three cases, the
distribution of strategies among the active agents is
completely random, with no particular correlation
among the strategies at the disposal of each agent. 
The initial score of the strategies is set to zero, and thus the $s$
strategies assigned to each agent are equally good.

As mentioned in the introduction, the existence of $N_f$ players
which always make the same choice is equivalent to a minority game
with no fixed players, but 
where the ``comfort'' threshold  has been shifted. 
We can assume
that there are only $N - N_f$ agents, but that the winning outcome is
0 when the number of agents which make that choice is less
than $N/2$, which, in this case, is greater than one half of
the number of agents.
\section{Results\label{s:results}}
\subsection{The phase transition.}
The analytical study of the minority game~\cite{CMZ00} allows us to
determine the existence and location,
of the transition between the ergodic and non
ergodic phase as function of the number of fixed agents, $N_f$.
Following ref.~\cite{CMZ00b}, the fixed agents considered
here play the role of  ``producers''.
These producers have only one strategy, which, in our case,
is the same for all of them, and it gives the same decision for each possible history,
so that, in the notation
in~\cite{CMZ00b}, they are maximally correlated.
The analytical results obtained there, valid in
the  ``thermodynamic''  limit when the number
of agents and available strategies is large, 
describe how the phase transition between the symmetric and
asymmetric phase depends on the 
the fraction  of effective producers, which
can be written as
$\rho \approx N_f^2 / N$. This number is  
large, $\rho \gg 1$, which allows
us to expand the implicit expression
for the critical value of $\rho$,
in the Appendix in~\cite{CMZ00b}. We find that, at the transition:
\begin{equation}
N_f^* \approx \frac{N}{2^{m/2} \sqrt{\pi}}
\label{transition}
\end{equation}
when the number of strategies per agent is $s=2$. When the number of
fixed agents, $N_f > N_f^*$, the game is in the ergodic phase.
The expression in eq.~(\ref{transition}) ceases
to be valid when the assumption $\rho \gg 1$ fails, that is, for
$N_f^* \ll \sqrt{N}$. 

When $N_f = 0$ and $s = 2$, there is a phase
transition for $N = N^* \approx 2^m / 0.3374$. 
\begin{figure}
\resizebox{8cm}{!}{\rotatebox{0}{\includegraphics{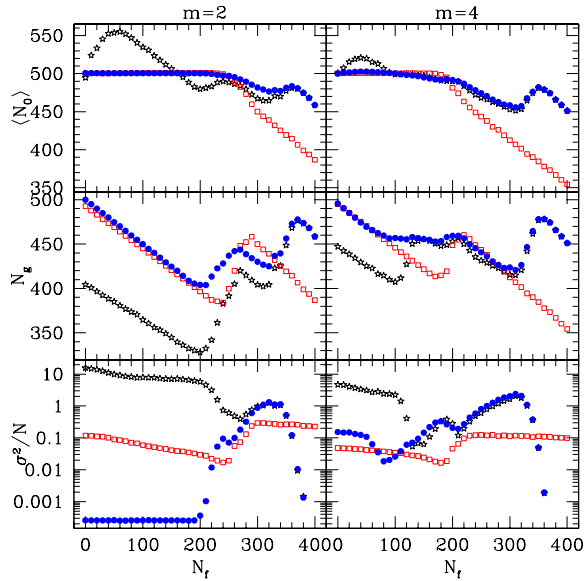}}}
\caption{Results for different quantities in the three versions of
the minority game discussed in the text, as function of
the number of agents which make the fixed choice 1. The total number
of agents is 1001. The number of strategies per agent
is $s=2$. Right column: $m=2$.
Left column: $m=4$. Stars: standard minority
game. Squares: Individual minority game. Solid circles: Minority game where
the impact of the strategy used is taken into account.
Top: Average number of agents which make choice 0. Middle: Average 
number of winners. Bottom: dispersion in the value of the number of agents
which make choice 0.}
\label{figure_1}
\end{figure}
\begin{figure}
\resizebox{8cm}{!}{\rotatebox{0}{\includegraphics{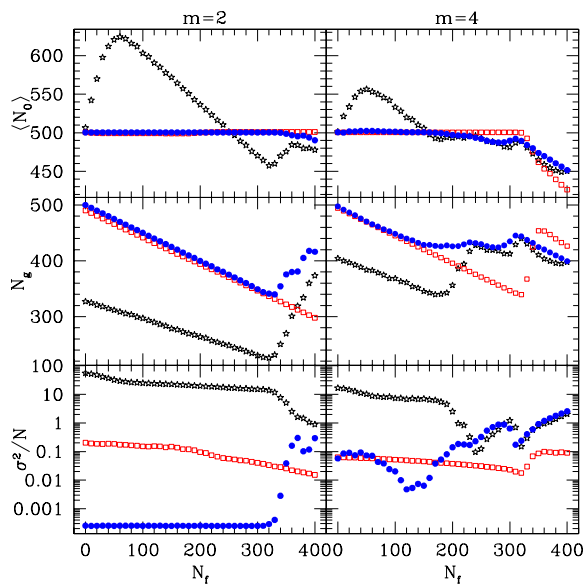}}}
\caption{As in Fig.~[\protect\ref{figure_1}], but for $s=4$.
Note the different scale used for the vertical axes.}
\label{figure_2}
\end{figure}
Results for different quantities and for the three versions of the minority
game described in the previous section are shown in Fig.~[\ref{figure_1}].
The number of agents is 1001. All quantities displayed
have been  calculated by averaging over 100 series of $300 \times
2^m$ time steps, after the system has achieved a stationary state,
each series corresponding to a different initial distribution
of strategies among the active agents. 

In the ergodic phase the discount of the market impact
ceases to be relevant, and the volatility
displayed in Fig.~[\ref{figure_1}] is the same for
cases i) and iii) defined in the previous section.
The value $N_f^*$ at which this transition takes place
is well described by eq.~(\ref{transition}).
Similar results are obtained for other values of the number
of strategies assigned to each agent, as shown in Fig.~[\ref{figure_2}].
The transition is shifted towards higher values of $N_f$ as we increase the number
of strategies $s$ available to each agent, and the
tendency for the agents to overreact to the information provided by the
fixed agents in the standard version of the game and in the non
ergodic phase is more pronounced (see $\langle N_0\rangle$ in figs.~[\ref{figure_1}]
and~[\ref{figure_2}]).

The fluctuations in the size
of the minority group are reduced by the presence of fixed agents,
in qualitative agreement with~\cite{Yetal03}. The fraction of winners
among the active agents is constant,
within our numerical accuracy, in the non ergodic phase,
and then it increases significantly as one enters the ergodic phase.
The increase is also in agreement with the results in~\cite{Yetal03},
where the efficiency of the game was increased with a biased
pool of strategies. The initial plateau of the fraction of
winners among the active players also
agrees qualitatively with the slow rise found in~\cite{Yetal03}.

\begin{figure}
\resizebox{8cm}{!}{\rotatebox{0}{\includegraphics{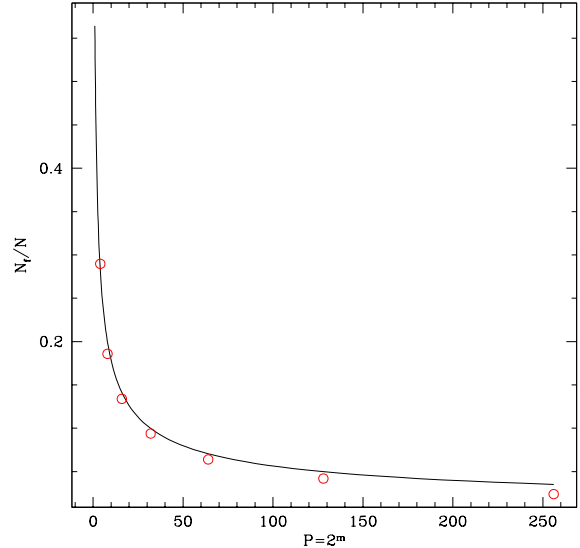}}}
\caption{Transition between the ergodic and the non ergodic phase
as function of the agents' memory, $m$, and $s=2$. Circles: numerical results.
Full line: $N_f / N = 1 / \sqrt{\pi P}$, using the
analysis in \protect~\cite{CMZ00b,CMO03}.}
\label{Nf_transition}
\end{figure}

The transition from the non ergodic (symmetric) phase to the ergodic
(asymmetric) phase, as function of
the memory of the agents, $m$, is shown in Fig.~[\ref{Nf_transition}].
For comparison, we also show the transition line obtained
using the analytical calculation in~\cite{CMZ00b}.

When the number of fixed agents is sufficiently large, the number of
winners is equal to the number of agents which make choice 0, that is,
which avoid the group chosen by the fixed agents. In addition, the 
dynamics converges to a stationary state where all agents which have
the appropriate strategies available make choice 0. There is only
one history describing the winning choice in this regime, $ ... 00000 ...$. 
On average, there is a fraction $2^{-m}$ of active agents which
cannot make use of the winning strategies, $ ... 00000 ...
\rightarrow 0$. Thus, the average number of active agents which make the
correct choice,  0,  is $N_0 = ( 1 - 2^{-m} ) ( N - N_f )$. 
This situation is stable if the $N_0$ agents are indeed in the minority
group, that is,
if $N_0 < N / 2$, which implies:
\begin{equation}
N_f > \frac{2^{m-1} - 1}{2^m - 1} N
\label{stationary}
\end{equation}
This inequality gives the threshold for the trivial dynamics when the number
of fixed agents is sufficiently large, deep into the ergodic phase.

\subsection{The non ergodic phase. ``Overscreening'' effects.}
Inside the non ergodic phase, $N_f < N_f^*$, where
the value of $N_f^*$ is given in eq.~(\ref{transition}),
the number of agents making the opposite choice of that of the fixed
agents, $N_0$, is such that $N_0 \ge N / 2$ in the non ergodic phase,
when the number of fixed agents is not zero (see fig.~[\ref{histories}]). This choice
makes the outcome highly unfavorable for the collective of active agents
as the number of agents which can win cannot exceed $N/2$.
\begin{figure}
\resizebox{8cm}{!}{\rotatebox{0}{\includegraphics{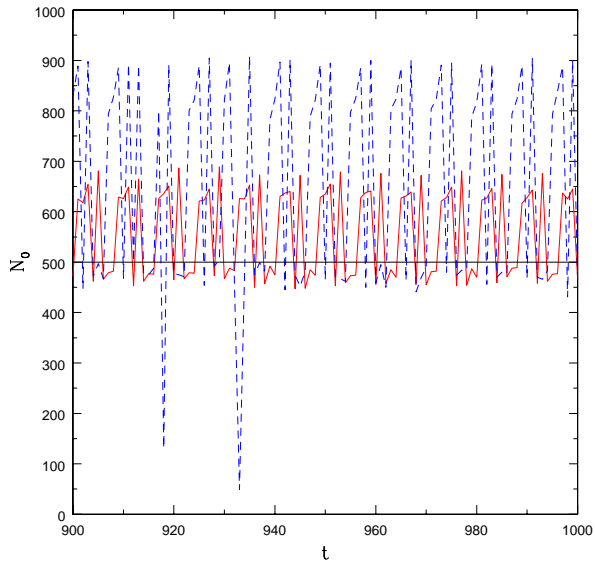}}}
\caption{
Number of agents which make choice 0 (the opposite to that made
by the fixed agents) as function of time, for a given initial 
distribution of strategies, and different
number of strategies, $s$, per agent.  The total number
of agents is $N = 1001$, the memory is $m= 2$,
and the number of fixed agents
is $N_f = 50$. Solid line: $s=2$.
Broken line: $s=6$.}
\label{time_series}
\end{figure}

The benefits of the collective in the non ergodic phase increase
greatly when the agents are able to discount the impact
of the strategies on the outcome, or use their own individual
histories. Note that when agents discount the impact of the
strategies, the score of a given strategy is not the same for
all agents, making the situation somewhat similar to that
in the individual minority game.  In both variations of the 
standard minority game the number of winning agents, $N_w$, is about
half the number of active agents, 
$N_w \approx ( N - N_f ) / 2$. This is below the maximum
number of possible winners, which is $N/2$, provided that $N_f < N/2$.

The most striking result is that, in the standard version of the
minority game, the number of active agents which make choice 0
is larger than its optimal value in the non ergodic phase.
This choice is
the opposite to the choice made by the fixed agents. Thus,
the active agents perceive the existence of the fixed
agents, but there is a herding effect which induces them to
make the opposite choice in numbers above the appropriate comfort level.
In the language of a random spin model~\cite{MPV87}, the active agents
``overscreen'' the external field induced by the fixed agents.
\subsection{The non ergodic phase. Dynamics.} 
We have analyzed the ``overscreening'' of the information provided
by the fixed agents in the standard minority
game by studying individual time series in the 
non ergodic phase. Results are shown in Fig.~[\ref{time_series}], for
a memory of two time steps, $m=2$, and $s=2$ and $s=6$ strategies per
active agent. The time series show well defined cycles with periodicity
greater than the time horizon available to the agents, who are
unable to make use of this information, as in the standard
minority game without fixed agents~\cite{SMR97}. These cycles are
sometimes interrupted by strong deviations when $s=6$. The origin of
these spikes is unclear, although it is consistent with the
enhancement of herding effects as the value of $s$ increases.

For the case $m=2$ the typical cycle spans eight time steps, where
the winning choice follows the series $... 11100010 ...$. We can 
understand this cycle by assuming that there are two
strategies with the highest score being used
by the agents. These strategies are mutually opposite. They can be
considered as representative of broad classes of strategies
with similar outcomes~\cite{CZ98}. We further assume that the score
of these strategies can either differ by one unit (the minimum
amount) or be equal, in which case the strategy used is decided
by a coin toss.
Then, i) At the beginning of the cycle the history processed
by the agents is $11$. We assume that
the strategy with the highest score predicts $11 \rightarrow 0$.
The majority of active agents follows this strategy so that the winning
choice is 1. The score of this strategy and that of its 
opposite, $11 \rightarrow 1$ becomes equal. ii) The history processed
by the agents remains $11$. The active agents take a random decision,
and the majority group is determined by the fixed agents.
The winning choice is 0, and the strategy $11 \rightarrow 0$
becomes again the one with the highest score. iii) We now assume that
the strategies predicting the two opposite outcomes after
the history $10$ have equal score. The active agents make a
random decision, and the outcome is determined by the
choice of the fixed agents. The winning choice is 0,
and the scores are updated accordingly. iv) We assume again
that the strategies $00 \rightarrow 0$ and $00 \rightarrow 1$ have the
same score. The winning choice is 0, and the scores
are updated. v) The active agents use the history $00$ and take choice 0.
The winning choice becomes 1, and the two strategies
$00 \rightarrow 0$ and $00 \rightarrow 1$ have again
the same score. vi) The history is $01$. If the strategies predicting
the two possible outcomes have the same score, the winning
choice will be 0. vii) The history now is $10$. The strategy with
the highest score is $10 \rightarrow 0$, as fixed in step iii).
The winning choice is 1. viii) The history is $01$. The strategy with
the highest score is $01 \rightarrow 0$, as fixed in step vi). The winning
choice is 1, and the cycle repeats itself. This succession of events is
schematically shown in Table[\ref{table}].
\begin{table}
\begin{tabular}{||c|c|c||}
\hline \hline
History &Strategy with the &Winning choice \\
&highest score & \\ \hline 
11 &$11 \rightarrow 0$ &1 \\ \hline 
11 &tie &0  \\ \hline 
10 &tie &0 \\ \hline 
00 &tie &0 \\ \hline 
00 &$00 \rightarrow 0$ &1 \\ \hline
01 &tie &0 \\ \hline 
10 &$10 \rightarrow 0$ &1 \\ \hline
01 &$01 \rightarrow 0$ &1  
\\  \hline  \hline
\end{tabular}
\caption{Histories and strategies which lead to the cycle
shown in Fig.~[\protect\ref{time_series}].}
\label{table}
\end{table}

The influence of the fixed agents is to determine the outcome
in cases where two strategies which lead to opposite choices
have the same score. The existence of these situations in the
standard minority game leads to a rich structure in the 
size of the groups~\cite{CPG99}, and to Gaussian fluctuations
around the average values, due to the randomness in the outcomes.
This randomness disappears in the presence of fixed agents. 
When two opposite strategies have the same score, around
half of the active agents make one choice and the other
half makes the opposite. The existence of fixed agents determines
the majority group, which is that chosen by the fixed agents.
Then, the active agents have a strong bias towards the opposite
group the next time that the same history presents itself. 
This tendency leads to the overscreening of the information 
provided by the fixed agents. If the number of time steps in the
history processed by the agents is $m$, the cycle is usually
of $2^{m+1}$ steps. These cycles appear in the non ergodic
phase, where a majority of active agents are able to 
distinguish the ``best'' strategies. The tendency towards
overscreening increases with the number $s$ of strategies
available to the agents. 

\begin{figure*}
\resizebox{0.7\textwidth}{!}{\rotatebox{0}{\includegraphics{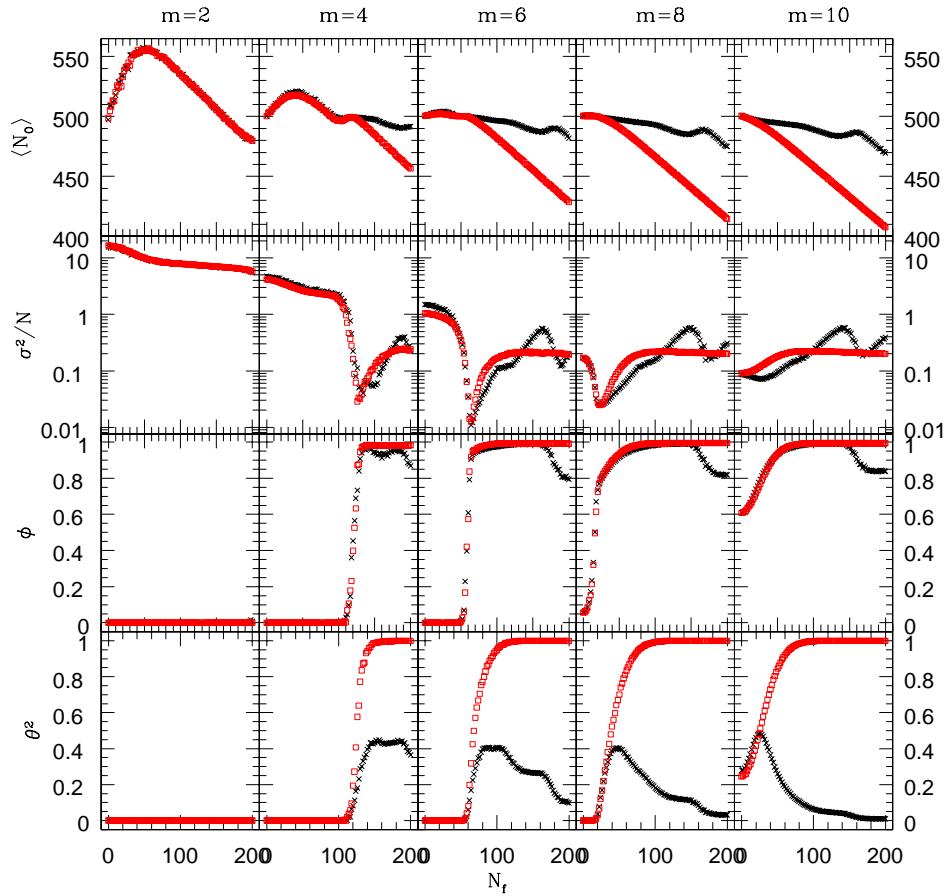}}}
\caption{Comparison of results for the standard minority
game with fixed agents obtained with the histories generated by
the dynamics of the game (crosses), and random histories (squares). First row:
Average number of agents which make the choice opposite to that
of the fixed agents, $\langle N_0 \rangle$. Second row:
Dispersion in $N_0$. Third row: Fraction of frozen agents.
Fourth row: Information stored in the dynamics (see
\protect\cite{CMZ00} ). The number of fixed
agents, $N_f$, is represented in the horizontal axes.
The number of strategies available to the agents is $s=2$.
Different columns correspond to different history lengths,
$m = 2,4,6,8,10$.
The results are calculated for a total of
$N = 1001$ agents, averaged over 100 initial distributions of
strategies and $300\times 2^m$ timesteps.}
\label{histories}
\end{figure*}

The information available  to the agents
in the series in Fig.~[\ref{time_series}],
vanishes, because the outcome after
a given history is totally unpredictable. Then, as
we are using a binary payoff, the difference between the scores
of the $s$ strategies of each agent averages to zero.
This can be seen in Fig.~[\ref{histories}], where we show how $\theta$
($\theta^2=\frac{1}{2^m}\sum_{\mu}\langle(2\psi-1)|\mu\rangle^2$,
$\psi$ is the minority group in each time step), which indicates the amount
of information on the time series is zero in the non ergodic
phase.
%

\subsection{Influence of the payoff function.}
\begin{figure}[ht!]
\resizebox{8cm}{!}{\rotatebox{0}{\includegraphics{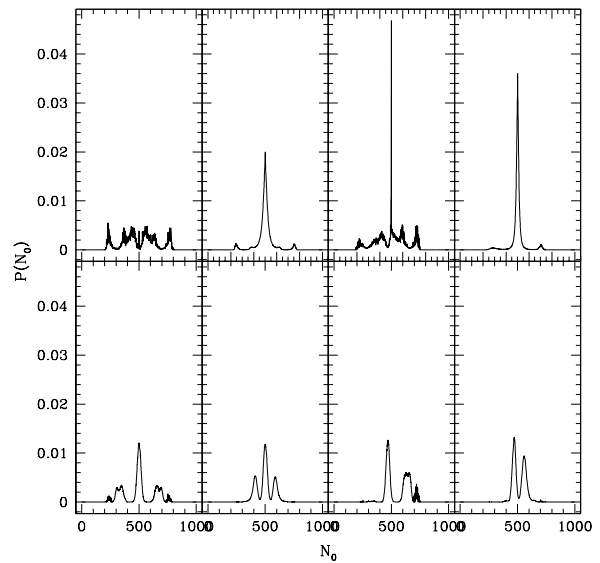}}}
\caption{Distribution $P ( N_0 )$ for the minority game with two
strategies per agent. The results are averaged
over 100 initial distributions of strategies.
The total number of
agents is $N = 1001$.
$N_f$ is the number of fixed agents.
Top: linear payoff. Left:  $m=2$ and $N_f = 0$.
Center left: $m=4$ and $N_f = 0$. Center right: $m=2$ and $N_f = 60$.
Right: $m=4$ and $N_f = 60$. 
Bottom: binary payoff. Left:  $m=2$ and $N_f = 0$.
Center left: $m=4$ and $N_f = 0$. Center right: $m=2$ and $N_f = 60$.
Right: $m=4$ and $N_f = 60$. }
\label{distribution}
\end{figure}

It is interesting to study the changes in the non ergodic phase
when the payoff function used in updating the score
of the strategies is proportional to
the deviation from the optimal occupancy,  instead of a step function,
as commonly used when analytical
methods are applied to the minority game~\cite{CMZ00}.
In this case, the condition that there is no information
available to the agents 
implies that the average payoff for each strategy
is zero. In order for this to happen, $\langle N_0
\rangle = N / 2$ should be satisfied. Hence, the tendency
towards overscreening in the ergodic phase described above 
does not lead to deviations of $\langle N_0 \rangle$ from its
``natural'' value. There is, however, a significant
asymmetry in the distribution $P ( N_0 )$, as shown
in Fig.~[\ref{distribution}]. There is a range of values of
$N_0$ near $N / 2$ for which $P ( N_0 )$ is biased towards
$N_0 > N / 2$, as in the minority game with a binary payoff.
This effect is compensated by the inverse asymmetry of
$P ( N_0 )$ when $N_0 \approx 0$ or $N_0 \approx N$.
Note that, for the standard situation with no fixed agents,
the distribution $P ( N_0 )$ is significantly different
from that in the minority game with a binary payoff~\cite{CPG99}. 

\subsection{Random vs. actual histories.}
We have studied the changes induced by replacing the actual
histories by random variables in the standard version
of the minority game, as initially
discussed in~\cite{C99}. The results are shown in Fig.~[\ref{histories}].
We plot there a number of statistical averages which serve
to characterize the minority game, like the fraction of
``frozen agents'', which although in principle active,
settle to use only one strategy at long times, $\phi$, and 
the information stored in the history of the game~\cite{CMZ00},
$\theta^2$, which is the bias towards one of the two outcomes of the game
when a given history appears on the time series\cite{outcome}.
In the non ergodic
phase, the agents are not able to distinguish between
the actual histories and a succession of random events,
as in the minority game without bias~\cite{C99}.
In the ergodic phase, however, the information provided
by the fixed agents becomes relevant, and there is
a difference between the actual game and that generated by
a succession of random events.

The cycles shown in Fig.~[\ref{time_series}] are due to the sequential 
substitution of the strategy with the highest score by its opposite.
Hence, a random succession of histories can give rise
to the same overscreening. Thus, the results in the non ergodic
phase do not change when the histories are random variables,
as shown in Fig.~[\ref{histories}].
The situation changes in the ergodic phase.
When the histories processed by the agents
are random but the winning choice is constant in time,
 0, the score of each strategy will depend only on how often
it contains  0 as an output. If the number of strategies at play
is small, in the ergodic phase,
the strategies with the highest score will contain
a significant number of 1's as outputs. Then, the agents
will become frozen, and make, with similar probability, the
two possible choices. This explains the results in the first
row and right columns (high values of
$m$) in Fig.~[\ref{histories}], where about half of the
active agents make the right choice, 0, and the other half choose 1, when
the histories are random. As shown in the third row, most agents
in this regime are frozen and
the information in the dynamically
generated histories is maximal~\cite{CMZ00}
(fourth row in Fig.~[\ref{histories}]).
This situation, where a significant number of active agents become frozen,
is more difficult to achieve when the histories are dynamically
generated by the agents themselves.

Finally, while the dynamics of the standard minority game in the non
ergodic phase imply
frequent situations where agents make choices using a coin toss,
see table[\ref{table}], the actual outcome of the game is
fixed. Hence, we do not expect that this non deterministic
aspect of the dynamics will play a significant role~\cite{JHJ02}. 
\section{Conclusions\label{s:conclusions}}
We have studied the minority game when a given number of agents 
make always the same choice. Hence, from the point of view of
the remaining, active, agents, the predictable behavior
of the fixed agents can be considered a source of
external information.
We have analyzed the standard version, the variant 
where agents use individual information, and that in which agents
are able to discount the effect of their actions on the outcome.

The system shows  a variety of interesting results in the non ergodic
phase, where the ability of the agents to process the information
available to them is highest:

i) In the standard minority game, the active agents tend to
overscreen the information provided by the fixed agents, leading
to disastrous effects for the collective (we are using the analogy with
spin models, where the external information can be viewed
as an applied field, to be screened by the dynamical spins
which represent the agents). The number of agents in
the minority group not only shows a large dispersion, as in
the symmetric minority game, but its average is far from
the optimal value. Assuming that there are $N$ agents, of which
$N_f$ make always the same choice, 1,
we find that the average of the number of
active agents which make choice 0 is $\langle N_0 \rangle > N / 2$, while
the optimal value is $\langle N_0 \rangle \approx N / 2$.

ii) The overscreening of the external information can be understood
through the existence of cycles longer than the amount of time steps
which the agents are able to process. The presence of the fixed
agents determine the outcome of the situations when opposite
strategies have the same score. This, in turn, leads to a strong bias
of the active agents towards the group not chosen by the fixed agents.
This bias proves catastrophic for the global benefit of the
active agents.

iii) The gain made by the collective of active agents in the non ergodic
phase is significantly improved when the agents use individual
information, or are able to discount the effect of their own choices
on the global outcome. The dispersion in the number of agents in
the minority group is greatly reduced. The average number of winners, $N_w$,
however, fluctuates around the half the number of active agents,
$\langle N_w \rangle \approx ( N - N_f ) / 2$, while the maximum
number of possible winners is $N / 2$, for $N_f < N / 2$. 

iv) The results for the non ergodic phase are qualitatively
the same when agents use individual information,
and for the case where agents discount the effect of their
strategies on the outcome.
This is probably due to the fact that, in both cases, the score assigned
to a given strategy is different for different agents.

These features imply that
the existence of an external bias in the non ergodic phase
of different versions of the minority game significantly reduces
the global benefit of the agents with respect
to the maximum possible value, which increases
as external information is fed into the system.

The situation where the agents
achieve the highest collective payoff, with respect to
the maximum payoff which can be achieved takes place when
the number of fixed agents vanishes. 
This fact is not contradictory with an increase in the
efficiency of the game in absolute terms~\cite{Yetal03}, as 
the opportunities for a given active agent are significantly
increased by the presence of fixed agents. 

On the other hand,
the predictability of the outcome, in the non ergodic phase, is zero,
as any outcome is possible after a given history~\cite{CMO03}, despite the fact
that the agents fail to guess the correct ``comfort''
threshold. 

We have not considered here the influence of
varying the initial scores of the strategies, which is
expected to change the volatility in the non ergodic phase~\cite{HC01,MC01}.
The fact that the active agents are unable to remove the external
information in the ergodic phase is in agreement with
the results in~\cite{CMO03}.
 
\section{Acknowledgements\label{s:acknowledgements}}
We appreciate helpful comments, and a careful reading of
the manuscript, to M. Marsili and D. Challet.
We are thankful to Caja de Ahorros de Granada ``La General''
and MCyT (Spain) for financial support, through grant
MAT2002-04095-C02-01.


\begin{thebibliography}{99}
\bibitem{note}
For completeness, the numbers of active agents making choices
0 and 1, $N_0$ and $N_1$
can be written using the notation in~\protect\cite{CMO03} as
$N_0 = N_a - A$, and $N_1 = N_a / 2 + A - L$, where $N_a$
is the number of active agents (denoted as $N$
in\protect\cite{CMO03}).
\bibitem{outcome}
Note that in the non ergodic phase the information is arbitraged away,
and the outcome of the game after a given history is unpredictable.
\end{thebibliography}
\end{document}